\documentclass[aps,prd,showpacs,nofootinbib]{revtex4}
\usepackage{bbm}
\usepackage{mathrsfs}
\usepackage{epsfig,psfrag}
\usepackage{amsmath,amsfonts,amssymb}
\usepackage[usenames]{color}
\usepackage{bm}
\usepackage{ulem}
\begin{document}

\title{Casimir Energies of Periodic Dielectric Gratings}
 
\author{Noah Graham}
\email{ngraham@middlebury.edu}
\affiliation{Department of Physics,
Middlebury College,
Middlebury, VT 05753  USA}

\begin{abstract}
Reflection of electromagnetic waves from a periodic grating can be
described in terms of a discrete coupled multichannel scattering
problem.  By modeling the grating as a space- and frequency-dependent
dielectric, it is possible to use a variable phase method, applied to
a generalized Helmholtz equation incorporating both transverse and
longitudinal modes, to efficiently compute the scattering $S$-matrix.
The projection onto transverse modes of this result, evaluated for
imaginary wave vector, provides the information necessary for a
Casimir energy calculation.  This approach is of particular interest
for gratings with deep corrugations, which can limit the applicability
of techniques based on the Rayleigh expansion.  We demonstrate the
method by calculating the Casimir interaction energy between
sinusoidal grating profiles as a function of separation and lateral
displacement.
\end{abstract}

\maketitle

\section{Introduction}

The Casimir force, arising from fluctuations of charges and fields in
quantum electrodynamics, has entered an era of unprecedented
precision measurements.  One particularly appealing application is
to the case of periodic gratings, where both lateral and
perpendicular forces can be measured 
\cite{PhysRevB.81.115417,Bao:2010zz,Chan:2011za,Intravaia,PhysRevB.89.235436}.
On the theoretical side, one can use Rayleigh expansion methods for
rectangular \cite{PhysRevLett.101.160403} and lamellar
\cite{Lussange:2012pta,Intravaia:2012iw,Guerout:2013ana} gratings, 
$C$ methods \cite{PhysRevA.90.012516}, and perturbative methods
\cite{Rodrigues:2006ku,Rodrigues:2007zza,Chen:2007jea} to calculate
Casimir interaction energies of periodic dielectrics or conductors.  These
approaches are often limited in their ability to handle deep corrugations
\cite{Rayleigh07,Lippman53,Uretsky65,Millar69,Millar71,Waterman74,Tatarskii95}
(for perfectly conducting rectangular corrugations, one can
obtain exact results using path integral techniques 
\cite{Emig2003,Buscher:2004tb}).  Here we take a different approach
and model the grating as a smooth dielectric function that depends on
$z$, the distance perpendicular to the grating, is periodic 
with period $L$ in the transverse direction $x$, and is independent of
the transverse direction $y$.  The dielectric function can also depend
on frequency, including dissipation (e.g. a Drude model).  We will use
scattering theory methods
\cite{Kats77,Jaekel91,Bulgac06,Lambrecht06,Kenneth06,spheres,Kenneth08,universal}
to find the Casimir energy for two such gratings in terms of the
reflection matrix for scattering from each grating individually.  We
compute these reflection matrices using a variable phase approach
\cite{variable,Graham09}, which enables us to solve the scattering
theory problem by integrating an ordinary matrix differential equation
in $z$ for the coupled modes arising from a Fourier decomposition in
$x$.  This calculation is the Cartesian analog of the approach in
Ref.~\cite{Forrow:2012sp}, which shows how to obtain scattering
matrices of asymmetric objects by solving coupled ordinary
differential equations in a spherical basis.  For the grating case,
the band structure in the periodic direction leads to a discrete
scattering problem, since mixing only takes place 
between modes whose wave numbers in the $x$ direction differ by an
integer multiple of $\frac{2\pi}{L}$.  Because the method
obtains the full $S$-matrix without approximation, it can be applied
to an arbitrary background profile, limited only by the computational
resources needed to describe the Fourier decomposition of the
dielectric background.

In the next section, we show how to compute the electromagnetic
reflection matrices for scattering from a periodic dielectric using the
generalized variable phase method.  Then, in the following section, we
demonstrate the method by using this scattering data to carry out a
sample calculation of the Casimir interaction energy of two dielectric
gratings.

\section{Scattering Calculation}

We formulate the problem as Maxwell scattering from a periodic,
frequency- and space-dependent dielectric $\epsilon(k,\bm{r})$,
\begin{equation}
\nabla \times \nabla \times \bm{E}_k(\bm{r}) = 
k^2 \epsilon(k, \bm{r}) \bm{E}_k(\bm{r}) \,,
\label{eqn:Maxwell}
\end{equation}
where $\omega = ck$ is the wave frequency.  This equation has
transverse solutions with $k\neq 0$, as well as unphysical
longitudinal solutions with $k=0$.  We use the technique of
Ref.~\cite{Forrow:2012sp}, in which we instead solve the generalized
Helmholtz equation
\begin{equation}
\nabla \times \nabla \times \bm{E}_k(\bm{r}) - \epsilon(k,\bm{r})
\nabla [\nabla \cdot (\epsilon(k,\bm{r}) \bm{E}_k(\bm{r}))]
 = k^2 \epsilon(k,\bm{r}) \bm{E}_k(\bm{r})\,.
\label{eqn:generalizedHelmholtz}
\end{equation}
Eq.~(\ref{eqn:generalizedHelmholtz}) shares the same transverse
solutions as Eq.~(\ref{eqn:Maxwell}), but it has longitudinal solutions
for which $k$ is related to wavelength and frequency in the usual way,
rather than having $k=0$.  Our method will include these longitudinal
modes in the calculation, so that we are working with a nonsingular
differential operator.  We then discard the longitudinal modes at the
end of the calculation via projection onto the subspace of transverse modes.

We consider the case where $\epsilon(k,\bm{r})$ is periodic in the $x$
direction with period $L$ and independent of $y$. (It is
straightforward to extend this formalism to the case where $\epsilon(k,
\bm{r})$ is periodic in both the $x$ and $y$ directions.)  As a result,
we can write $\epsilon(k,\bm{r})$ as a Fourier series,
\begin{equation}
\epsilon(k,\bm{r}) = \sum_{n=-\infty}^\infty \epsilon_{n}(k,z) e^{2 \pi
i n x/L} \,,
\end{equation}
where $\epsilon_{n}(k,z) \to \delta_{n0}$ for $z\to\pm \infty$.
For simplicity, we also assume that $\epsilon(k,\bm{r})$ is symmetric in
$z$.  We can then write the general solution to
Eq.~(\ref{eqn:generalizedHelmholtz}) as
\begin{equation}
\bm{E}(\bm{r}) = \int_{-\infty}^\infty \frac{dk_x}{2\pi}
\int_{-\infty}^\infty \frac{dk_y}{2\pi}
\sum_{i,\chi} E_{i,\chi,k_x,k_y}(k,z) \hat{\bm{e}}_i e^{i k_x x}
e^{i k_y y} \,,
\label{eqn:generalsol}
\end{equation}
where we have decomposed the solutions in terms of the transverse
momenta $k_x$ and $k_y$, the vector component $i=x,y,z$, and the
parity under $z$ reflection $\chi =\pm 1$.

Because the dielectric background is periodic in $x$ and independent
of $y$, the only values of $k_x$ that mix are those differing by
integer multiples of $\frac{2\pi}{L}$, while the $k_y$ values do not
mix at all.  The result is a band structure in $k_x$ where each
scattering channel is labeled by $k_{x_0}$ ranging from 
$-\frac{\pi}{L}$ to $\frac{\pi}{L}$ and comprises a discrete set of 
$k_x$ values, differing from $k_{x_0}$ by an integer times 
$\frac{2\pi}{L}$.
We therefore solve a matrix scattering problem for modes
with $k_x = k_{x_0} + \frac{2\pi n}{L}$, indexed by
the integer $n$, and a fixed value of $k_y=k_{y_0}$.  We define
$k_z =k \sqrt{1 - \frac{k_x^2 + k_y^2}{k^2}}$, and each mode has three
spatial components since we have a vector field.  For a given $k$ and
$\chi$, we obtain a separate matrix scattering problem for each value of
$k_{x_0}$ between $-\frac{\pi}{L}$ and $\frac{\pi}{L}$ and each value
of $k_{y_0}$ from zero to infinity.  On the real $k$-axis, this matrix
is finite-dimensional, since only modes with $|n| <
\frac{L}{2\pi}\left(|k| - \sqrt{k_{x_0}^2 + k_{y_0}^2}\right)$, those
that represent propagating rather than evanescent waves, contribute to
the $S$-matrix.  To compute the Casimir energy, we will analytically
continue to the imaginary $k$-axis, in which case there will be no
limit on $|n|$; as usual, we will be able to truncate these matrices
at large wave number for numerical calculations.  Even though the
ultimate calculation will be carried out on the imaginary axis, it is
helpful to begin from the real axis, because there the unitarity of
the finite-dimensional $S$-matrix gives a strong check of the
numerical calculation.

In the region where the dielectric differs from vacuum, the
decomposition of scattering solutions into transverse and longitudinal
components is highly nontrivial.  The $S$-matrix, however, is defined in
terms of free asymptotic waves, for which it is straightforward to
identify transverse and longitudinal modes.  We have the
TE and TM transverse modes respectively,
\begin{equation}
\bm{M}(k,k_x,k_y,\bm{r}) =
\frac{k_y \hat{\bm{x}}-k_x \hat{\bm{y}}}{\sqrt{k_x^2 + k_y^2}} 
e^{i\bm{k}\cdot\bm{r}}
\hbox{\qquad and \qquad}
\bm{N}(k,k_x,k_y,\bm{r}) =
\frac{k_z \hat{\bm{k}} - k \hat{\bm{z}}}{\sqrt{k_x^2 + k_y^2}}
e^{i\bm{k}\cdot\bm{r}} \,,
\label{eqn:transverse}
\end{equation}
and the longitudinal mode
\begin{equation}
\bm{L}(k,k_x,k_y,\bm{r}) =
\frac{\bm{k}}{k} e^{i\bm{k}\cdot\bm{r}} \,,
\end{equation}
where $\bm{k} = (k_x,k_y,k_z)$ and $k=|\bm{k}|$.
The $S$-matrix for scattering governed by
Eq.~(\ref{eqn:generalizedHelmholtz}) must commute with the projection
onto the subspace of transverse modes (for both real and imaginary wave
number), which gives another strong check on our numerical calculations.

To define the $S$-matrix, we combine the solutions with $k$ and
$-k$ (or, equivalently, the outgoing wave solution and its
conjugate, the incoming wave solution) to form the physical wave
functions \cite{Newton02} in the symmetric and antisymmetric channels
under reflection in $z$,
\begin{equation}
\hat \psi_{\pm}(k,k_{x_0},k_{y_0},z) = \pm \hat F(-k,k_{x_0},k_{y_0},z)
\hat M + \hat F(k,k_{x_0},k_{y_0},z) \hat
S_{\pm}(k,k_{x_0},k_{y_0})
\end{equation}
respectively, where $F(k,k_{x_0},k_{y_0},z)$ is the
outgoing wave solution, written as a matrix in the vectorspace of
modes with $k_x = k_{x_0} + \frac{2 \pi n}{L}$ and $k_y=k_{y_0}$.  
Here $\hat M$ is a diagonal matrix with $+1$ on the
diagonal for $x$ and $y$ vector components and $-1$ for $z$
components.  This matrix captures the additional minus sign involved
in imposing the parity boundary conditions on the $z$-component of the
vector wave function at $z=0$.  On the real axis, the resulting $\hat
S$ matrix is unitary and finite-dimensional: it only involves
asymptotic propagating waves with $|n| < \frac{L}{2\pi}\left(|k| -
\sqrt{k_{x_0}^2 +  k_{y_0}^2}\right)$.

To compute the $S$-matrix, we consider both the regular and outgoing
solutions to the generalized Helmholtz equation in
Eq.~(\ref{eqn:generalizedHelmholtz}).  In doing so, it will be helpful
to parameterize these solutions in a way that factors out the free
solutions \cite{density}.  Defining $\hat k_z(k,k_{x_0},k_{y_0})$ to
be a diagonal matrix with $k_z$ on the diagonal, we write the outgoing
solution as
\begin{equation}
\hat F(k,k_{x_0},k_{y_0},z)  = \hat G(k,k_{x_0},k_{y_0},z) 
\exp[i \hat k_z(k,k_{x_0},k_{y_0})z]
\end{equation}
and the transpose of the regular solution as (note the reversed order)
\begin{equation}
\hat \Phi_{\pm}(k,k_{x_0},k_{y_0},z)^t  = 
\exp[\pm i \hat M \hat k_z(k,k_{x_0},k_{y_0})z]
H_{\pm}(k,k_{x_0},k_{y_0},z) \,.
\end{equation}
The regular solution is different in the two parity channels, but
the outgoing solution is the same.  Plugging these solutions into
Eq.~(\ref{eqn:generalizedHelmholtz}), we obtain equations of the form
\cite{Forrow:2012sp}
\begin{eqnarray}
0 = -\frac{d^2}{dz^2}\hat G(k,k_{x_0},k_{y_0},z) + 
\left(\hat D_1(k,k_{x_0},k_{y_0},z) \hat G(k,k_{x_0},k_{y_0},z)
- 2 \frac{d}{dz}\hat G(k,k_{x_0},k_{y_0},z)\right)
i \hat k_z(k,k_{x_0},k_{y_0}) \cr 
+ \hat D_1(k,k_{x_0},k_{y_0},z) \frac{d}{dz}\hat G(k,k_{x_0},k_{y_0},z)
+ \hat D_0(k,k_{x_0},k_{y_0},z)\hat G(k,k_{x_0},k_{y_0},z) + 
\hat G(k,k_{x_0},k_{y_0},z)\hat k_z(k,k_{x_0},k_{y_0})^2
\label{eqn:g}
\end{eqnarray}
and
\begin{eqnarray}
0=-\frac{d^2}{dz^2}\hat H_\pm(k,k_{x_0},k_{y_0},z) 
\mp i\hat M \hat k_z (k,k_{x_0},k_{y_0})
\left(\hat H_\pm(k,k_{x_0},k_{y_0},z) \hat D_1(k,k_{x_0},k_{y_0},z) +
2\frac{d}{dz} \hat H_\pm(k,k_{x_0},k_{y_0},z)\right) \cr
- \left(\frac{d}{dz}\hat H_\pm(k,k_{x_0},k_{y_0},z)\right)
\hat D_1(k,k_{x_0},k_{y_0},z)
- \hat H_\pm(k,k_{x_0},k_{y_0},z)
\frac{d}{dz}\hat D_1(k,k_{x_0},k_{y_0},z) \cr
+ \hat H_\pm(k,k_{x_0},k_{y_0},z) \hat D_0(k,k_{x_0},k_{y_0},z)
+ \hat k_z(k,k_{x_0},k_{y_0})^2 H_\pm(k,k_{x_0},k_{y_0},z) \,, 
\end{eqnarray}
which are to be solved subject to the boundary conditions
\begin{equation}
\hat G(k,k_{x_0},k_{y_0},z\to\infty) = \hat 1 \qquad
\left.\frac{d}{dz} \hat G(k,k_{x_0},k_{y_0},z)\right\vert_{z\to\infty}
= \hat 0
\end{equation}
and
\begin{equation}
\hat H_{\pm}(k,k_{x_0},k_{y_0},z=0) = 
\hat h_{\pm}(k,k_{x_0},k_{y_0}) \qquad
\left.\frac{d}{dz} \hat H_{\pm}(k,k_{x_0},k_{y_0},z)\right\vert_{z=0}
= \hat 1\,,
\end{equation}
where $\hat h_+(k,k_{x_0},k_{y_0})$ is a diagonal matrix 
whose diagonal entries are zero for the $z$ components of
$E_{i,\chi,k_x,k_y}(z)$ and $(-ik_z)^{-1}$ for the $x$ and $y$
components, while 
$\hat h_-(k,k_{x_0},k_{y_0})$ is a diagonal matrix
whose diagonal entries are zero for the $x$ and $y$ components of
$E_{i,\chi,k_x,k_y}(z)$ and $(-ik_z)^{-1}$ for the $z$ components.
The matrices $\hat D_0(k,k_{x_0},k_{y_0},z)$ and
$\hat D_1(k,k_{x_0},k_{y_0},z)$ are the result of applying the vector
derivatives in Eq.~(\ref{eqn:generalizedHelmholtz}) to the general
solution in Eq.~(\ref{eqn:generalsol}) and
depend on the dielectric Fourier components $\epsilon_n(k,z)$ and
their derivatives.  These matrices are typically obtained from
symbolic computation, as shown in a sample calculation available from
{\tt http://community.middlebury.edu/\~{}ngraham}\,.

The $S$-matrix is then given by \cite{Newton02,Forrow:2012sp}
\begin{equation}
S_\pm(k,k_{x_0},k_{y_0}) = 
\widetilde{\cal W}_\pm(k,k_{x_0},k_{y_0})^{-1} \hat M 
\widetilde{\cal W}_\pm(-k,k_{x_0},k_{y_0}) \hat M 
\end{equation}
where $\widetilde{\cal W}_\pm(k,k_{x_0},k_{y_0})$
is the generalized Wronskian of the incoming and outgoing solutions,
\begin{eqnarray}
\left.\widetilde{\cal W}_\pm(k,k_{x_0},k_{y_0})\right\vert_z
&=& \left[\hat \Phi_\pm(k,k_{x_0},k_{y_0},z)^t \frac{d}{dz}\left(\hat
F(k,k_{x_0},k_{y_0},z) \right)
- \frac{d}{dz}\left(\hat \Phi_\pm(k,k_{x_0},k_{y_0},z)^t \right)
\hat F(k,k_{x_0},k_{y_0},z)  \right. \cr 
&& \quad - \left. \Phi_\pm(k,k_{x_0},k_{y_0},z)^t
\hat D_1(k,k_{x_0},k_{y_0},z) 
\hat F(k,k_{x_0},k_{y_0},z) \right] \hat N(k,k_{x_0},k_{y_0})\,,
\label{eqn:Wtilde}
\end{eqnarray}
which is independent of $z$.  Here $\hat N(k,k_{x_0},k_{y_0})$ is a
diagonal matrix with $\sqrt{\frac{k}{k_z}}$ on the diagonal, which
normalizes the incident flux in the different components of our
scattering basis, yielding a unitary $S$-matrix for real $k$.
Since the Wronskian can be evaluated at any value of $z$, in order to
optimize the numerical calculation, we will choose 
a common fitting point at the characteristic width of the potential,
and then evaluate the Wronskian by integrating $\hat G$ inward to this
point from infinity and $\hat H_\pm$ outward to this point from the
origin.  The numerical advantages of this approach become particularly
important for imaginary $k$, as will be required for our Casimir energy
calculation.  Although in principle the $S$-matrix could be obtained
from either $\hat G$ or $\hat H_\pm$ alone, this combined approach is
necessary on the imaginary $k$-axis to avoid numerical problems from
growing exponentials in that case.

With the $S$-matrix in hand, we are prepared to consider the Casimir
energy.  For two gratings whose origins are separated
by the vector $\Delta \bm{r}$, we define the translation matrix
$\hat U(k,k_{x_0},k_{y_0})$ as a diagonal matrix with
$\exp[i \bm{k} \cdot \Delta \bm{r}]$ on the diagonal.
For each grating, we form the reflection coefficient as the difference
between the $S$-matrices for the symmetric and antisymmetric channels,
\begin{equation}
\hat r(k,k_{x_0},k_{y_0}) = \frac{1}{2}\left[\hat S_+(k,k_{x_0},k_{y_0}) -
\hat S_-(k,k_{x_0},k_{y_0}) \right]\,.
\end{equation}
We then project both $\hat U(k,k_{x_0},k_{y_0})$ and
$\hat r(k,k_{x_0},k_{y_0})$ onto the subspace spanned by the
transverse modes in Eq.~(\ref{eqn:transverse}), denoting the results
as $\bar U(k,k_{x_0},k_{y_0})$ and $\bar r(k,k_{x_0},k_{y_0})$
respectively.  This projection discards the longitudinal modes,
leaving the transverse modes unchanged.

The Casimir interaction energy per unit area for two identical
gratings is then given by the scattering theory approach as
\cite{Kats77,Jaekel91,Lambrecht06,Kenneth06,spheres,Kenneth08,universal}
\begin{equation}
\frac{\cal E}{A} = \frac{\hbar c}{4\pi^3}
\int_{0}^\infty d\kappa
\int_{-\pi/L}^{\pi/L} dk_{x_0} \int_{0}^{\infty} dk_{y_0}
\log \det \left[1 - \bar U(i\kappa,k_{x_0},k_{y_0}) \bar
r(i\kappa,k_{x_0},k_{y_0})
\bar U(i\kappa,-k_{x_0},-k_{y_0}) \bar r(i\kappa,k_{x_0},k_{y_0})
\right] \,.
\label{eqn:energy}
\end{equation}
This result is now suitable for numerical computation.

\section{Applications and Discussion}

\begin{figure}[htbp]
\includegraphics[width=0.5\linewidth]{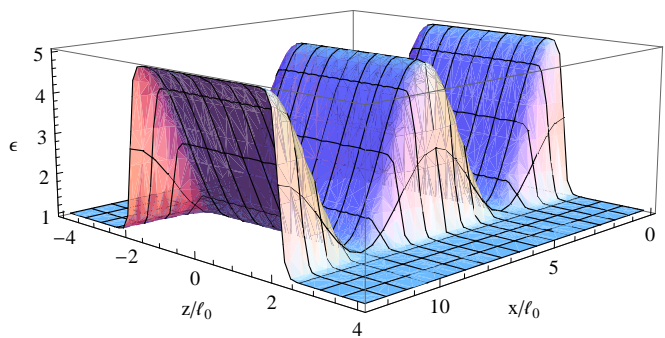}
\caption{(Color online)
Graph of the dielectric grating profile, as a function of $x$
and $z$.}
\label{fig:gratingplot}
\end{figure}

We illustrate the method by calculating the Casimir force between two
identical gratings with a frequency-independent dielectric function
given in terms of Fourier components
\begin{equation}
\epsilon_{0}(k,z) = 2 \epsilon_{1}(k,z) = 2 \epsilon_{-1}(k,z) = 
\frac{h}{1+\exp[s(|z|-w)]}
\end{equation}
with all other $\epsilon_{n}(k,z)=0$.  This profile gives a step
function shape in $z$; we choose height $h=2$, width $w=2 \ell_0$,
steepness $s=16/\ell_0$, and period $L=2\pi \ell_0$, where we work in
units of the length $\ell_0$.  The total
dielectric function is shown in Fig.~\ref{fig:gratingplot}.
We let $\Delta z$ be the separation between the gratings,
measured between their center planes, and $\Delta x$ be their
transverse displacement.

\begin{figure}[htbp]
\includegraphics[width=0.5\linewidth]{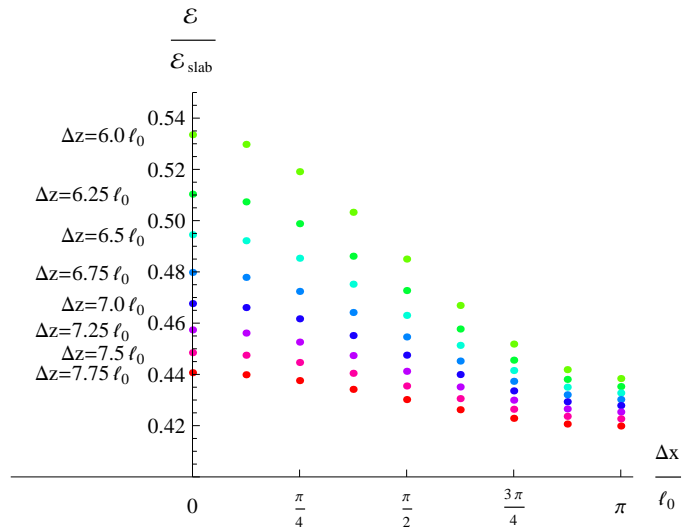}
\caption{(Color online)
Ratio of the Casimir energy of two dielectric gratings to the
Casimir energy for two slabs with the same maximum width and dielectric
constant, as a function of the separation $\Delta z$ and the lateral
displacement $\Delta x$.  As expected, the ratio is less than one, and
maximized when the two gratings are aligned in phase.}
\label{fig:gratingvectorcasimirplot}
\end{figure}

To see the effects of corrugations on the Casimir force, we compare
the Casimir energy of the two gratings to the Casimir energy of two
planar dielectric slabs with the same transverse area, maximum width
$2w$, and dielectric constant $\epsilon = 2h$.  For a slab, the
reflection matrix is diagonal, with diagonal entries given by the
Fresnel result
\begin{equation}
r_{\pm,\hbox{\tiny slab}} = 
\pm \frac{\Gamma_\pm (1-e^{4 i \beta w})}{1-\Gamma_\pm^2 e^{4 i \beta w}}
e^{-2i k_z w} \,,
\end{equation}
for the TE and TM polarization modes respectively, where
\begin{equation}
\Gamma_\pm = 
\frac{\cos \phi_i - \epsilon^{\pm \frac{1}{2}} \cos \phi_t}
{\cos \phi_i + \epsilon^{\pm \frac{1}{2}}\cos \phi_t} \,,
\end{equation}
the incident and transmitted angles are given by
\begin{equation}
\cos \phi_i = \frac{k_z}{k} \hbox{\qquad and \qquad}
\sin \phi_t = \frac{\sin \phi_i}{\sqrt{\epsilon}} \,,
\end{equation}
and $\beta = k \sqrt{\epsilon} \cos \phi_t$.  Results for the ratio of
Casimir energies $\frac{{\cal E}}{{\cal E}_{\hbox{\tiny slab}}}$ are
shown in Fig.~\ref{fig:gratingvectorcasimirplot}.  In this
calculation, the $\kappa$ and $k_{y_0}$ integrals in
Eq.~(\ref{eqn:energy}) are truncated at 
$k_{\hbox{\tiny min}} = 0.0078125/\ell_0$ and 
$k_{\hbox{\tiny max}} = 2.5/\ell_0$, and the
matrices in $k_x$ are truncated to include only modes within the same
$k_{\hbox{\tiny max}}$ of $k_{x_0}$.  For each $\kappa$, $k_{x_0}$,
and $k_{y_0}$, for both $\kappa$ and $-\kappa$ we integrate $\hat G$
and $\hat H_\pm$ from $z=0$ and $z=4w$ respectively to a common
fitting point at $z=w$.  We form the Wronskian in the symmetric and
antisymmetric channels to obtain the $S$-matrices in each channel,
which we then combine to form the reflection matrix.  Because of the
symmetry of our dielectric profile, the integrand is even in $k_{x_0}$
and we only need to compute the case of positive $k_{x_0}$.

As we expect, Fig.~\ref{fig:gratingvectorcasimirplot} shows that
the energy of the two gratings is smaller in magnitude than in the
case of two slabs, since we have effectively removed dielectric
material from the slabs to form the gratings.  The
dependence of the energy of the gratings on $\Delta x$ leads to a
lateral force trying to align the gratings (the energy for
the case of two slabs is negative and independent of $\Delta x$).  The
dependence on $\Delta x$ becomes weaker as $\Delta z$ increases, since
the calculation is increasingly dominated by long wavelength
fluctuations, which are less sensitive to the contours of the grating.

Having demonstrated the effectiveness of this method with a sample
calculation, a natural next step is to develop large-scale numerical
calculations, which can allow for closer separations by handling a
larger range of $k$ values, and can also include more Fourier
components in the dielectric function in order to focus on steep
gratings, for which calculations based on the Rayleigh expansion
are often not applicable.  Work in this direction is in progress
\cite{NGJSD}.

\begin{acknowledgments}
It is a pleasure to thank G.\ Bimonte, J.\ S.\ Dunham, T.\ Emig, R.\
L.\ Jaffe, M.\ Kardar, and M.\ Kr\"uger for helpful conversations.
N.\ G.\ was supported in part by the National Science Foundation (NSF)
through grant PHY-1213456.
\end{acknowledgments}

\bibliographystyle{apsrev}
\bibliography{grating}

\begin{thebibliography}{36}
\expandafter\ifx\csname natexlab\endcsname\relax\def\natexlab#1{#1}\fi
\expandafter\ifx\csname bibnamefont\endcsname\relax
  \def\bibnamefont#1{#1}\fi
\expandafter\ifx\csname bibfnamefont\endcsname\relax
  \def\bibfnamefont#1{#1}\fi
\expandafter\ifx\csname citenamefont\endcsname\relax
  \def\citenamefont#1{#1}\fi
\expandafter\ifx\csname url\endcsname\relax
  \def\url#1{\texttt{#1}}\fi
\expandafter\ifx\csname urlprefix\endcsname\relax\def\urlprefix{URL }\fi
\providecommand{\bibinfo}[2]{#2}
\providecommand{\eprint}[2][]{\url{#2}}

\bibitem[{\citenamefont{Chiu et~al.}(2010)\citenamefont{Chiu, Klimchitskaya,
  Marachevsky, Mostepanenko, and Mohideen}}]{PhysRevB.81.115417}
\bibinfo{author}{\bibfnamefont{H.-C.} \bibnamefont{Chiu}},
  \bibinfo{author}{\bibfnamefont{G.~L.} \bibnamefont{Klimchitskaya}},
  \bibinfo{author}{\bibfnamefont{V.~N.} \bibnamefont{Marachevsky}},
  \bibinfo{author}{\bibfnamefont{V.~M.} \bibnamefont{Mostepanenko}},
  \bibnamefont{and} \bibinfo{author}{\bibfnamefont{U.}~\bibnamefont{Mohideen}},
  \bibinfo{journal}{Phys. Rev. B} \textbf{\bibinfo{volume}{81}},
  \bibinfo{pages}{115417} (\bibinfo{year}{2010}).

\bibitem[{\citenamefont{Bao et~al.}(2010)\citenamefont{Bao, Guerout, Lussange,
  Lambrecht, Cirelli et~al.}}]{Bao:2010zz}
\bibinfo{author}{\bibfnamefont{Y.}~\bibnamefont{Bao}},
  \bibinfo{author}{\bibfnamefont{R.}~\bibnamefont{Guerout}},
  \bibinfo{author}{\bibfnamefont{J.}~\bibnamefont{Lussange}},
  \bibinfo{author}{\bibfnamefont{A.}~\bibnamefont{Lambrecht}},
  \bibinfo{author}{\bibfnamefont{R.}~\bibnamefont{Cirelli}},
  \bibnamefont{et~al.}, \bibinfo{journal}{Phys. Rev. Lett.}
  \textbf{\bibinfo{volume}{105}}, \bibinfo{pages}{250402}
  (\bibinfo{year}{2010}).

\bibitem[{\citenamefont{Chan et~al.}(2008)\citenamefont{Chan, Bao, Zou,
  Cirelli, Klemens et~al.}}]{Chan:2011za}
\bibinfo{author}{\bibfnamefont{H.}~\bibnamefont{Chan}},
  \bibinfo{author}{\bibfnamefont{Y.}~\bibnamefont{Bao}},
  \bibinfo{author}{\bibfnamefont{J.}~\bibnamefont{Zou}},
  \bibinfo{author}{\bibfnamefont{R.}~\bibnamefont{Cirelli}},
  \bibinfo{author}{\bibfnamefont{F.}~\bibnamefont{Klemens}},
  \bibnamefont{et~al.}, \bibinfo{journal}{Phys. Rev. Lett.}
  \textbf{\bibinfo{volume}{101}}, \bibinfo{pages}{030401}
  (\bibinfo{year}{2008}).

\bibitem[{\citenamefont{Intravaia et~al.}(2013)\citenamefont{Intravaia, Koev,
  Jung, Talin, Davids, Decca, Aksyuk, Dalvit, and Lopez}}]{Intravaia}
\bibinfo{author}{\bibfnamefont{F.}~\bibnamefont{Intravaia}},
  \bibinfo{author}{\bibfnamefont{S.}~\bibnamefont{Koev}},
  \bibinfo{author}{\bibfnamefont{I.~W.} \bibnamefont{Jung}},
  \bibinfo{author}{\bibfnamefont{A.~A.} \bibnamefont{Talin}},
  \bibinfo{author}{\bibfnamefont{P.~S.} \bibnamefont{Davids}},
  \bibinfo{author}{\bibfnamefont{R.~S.} \bibnamefont{Decca}},
  \bibinfo{author}{\bibfnamefont{V.~A.} \bibnamefont{Aksyuk}},
  \bibinfo{author}{\bibfnamefont{D.~A.~R.} \bibnamefont{Dalvit}},
  \bibnamefont{and} \bibinfo{author}{\bibfnamefont{D.}~\bibnamefont{Lopez}},
  \bibinfo{journal}{Nature Communications} \textbf{\bibinfo{volume}{4}},
  \bibinfo{pages}{2515} (\bibinfo{year}{2013}).

\bibitem[{\citenamefont{Banishev et~al.}(2014)\citenamefont{Banishev, Wagner,
  Emig, Zandi, and Mohideen}}]{PhysRevB.89.235436}
\bibinfo{author}{\bibfnamefont{A.~A.} \bibnamefont{Banishev}},
  \bibinfo{author}{\bibfnamefont{J.}~\bibnamefont{Wagner}},
  \bibinfo{author}{\bibfnamefont{T.}~\bibnamefont{Emig}},
  \bibinfo{author}{\bibfnamefont{R.}~\bibnamefont{Zandi}}, \bibnamefont{and}
  \bibinfo{author}{\bibfnamefont{U.}~\bibnamefont{Mohideen}},
  \bibinfo{journal}{Phys. Rev. B} \textbf{\bibinfo{volume}{89}},
  \bibinfo{pages}{235436} (\bibinfo{year}{2014}).

\bibitem[{\citenamefont{Lambrecht and
  Marachevsky}(2008)}]{PhysRevLett.101.160403}
\bibinfo{author}{\bibfnamefont{A.}~\bibnamefont{Lambrecht}} \bibnamefont{and}
  \bibinfo{author}{\bibfnamefont{V.~N.} \bibnamefont{Marachevsky}},
  \bibinfo{journal}{Phys. Rev. Lett.} \textbf{\bibinfo{volume}{101}},
  \bibinfo{pages}{160403} (\bibinfo{year}{2008}).

\bibitem[{\citenamefont{Lussange et~al.}(2012)\citenamefont{Lussange, Guerout,
  and Lambrecht}}]{Lussange:2012pta}
\bibinfo{author}{\bibfnamefont{J.}~\bibnamefont{Lussange}},
  \bibinfo{author}{\bibfnamefont{R.}~\bibnamefont{Guerout}}, \bibnamefont{and}
  \bibinfo{author}{\bibfnamefont{A.}~\bibnamefont{Lambrecht}},
  \bibinfo{journal}{Phys. Rev.} \textbf{\bibinfo{volume}{A86}},
  \bibinfo{pages}{062502} (\bibinfo{year}{2012}).

\bibitem[{\citenamefont{Intravaia et~al.}(2012)\citenamefont{Intravaia, Davids,
  Decca, Aksyuk, Lopez et~al.}}]{Intravaia:2012iw}
\bibinfo{author}{\bibfnamefont{F.}~\bibnamefont{Intravaia}},
  \bibinfo{author}{\bibfnamefont{P.}~\bibnamefont{Davids}},
  \bibinfo{author}{\bibfnamefont{R.}~\bibnamefont{Decca}},
  \bibinfo{author}{\bibfnamefont{V.}~\bibnamefont{Aksyuk}},
  \bibinfo{author}{\bibfnamefont{D.}~\bibnamefont{Lopez}},
  \bibnamefont{et~al.}, \bibinfo{journal}{Phys. Rev.}
  \textbf{\bibinfo{volume}{A86}}, \bibinfo{pages}{042101}
  (\bibinfo{year}{2012}).

\bibitem[{\citenamefont{Gu\'erout et~al.}(2013)\citenamefont{Gu\'erout,
  Lussange, Chan, Lambrecht, and Reynaud}}]{Guerout:2013ana}
\bibinfo{author}{\bibfnamefont{R.}~\bibnamefont{Gu\'erout}},
  \bibinfo{author}{\bibfnamefont{J.}~\bibnamefont{Lussange}},
  \bibinfo{author}{\bibfnamefont{H.}~\bibnamefont{Chan}},
  \bibinfo{author}{\bibfnamefont{A.}~\bibnamefont{Lambrecht}},
  \bibnamefont{and} \bibinfo{author}{\bibfnamefont{S.}~\bibnamefont{Reynaud}},
  \bibinfo{journal}{Phys. Rev.} \textbf{\bibinfo{volume}{A87}},
  \bibinfo{pages}{052514} (\bibinfo{year}{2013}).

\bibitem[{\citenamefont{Wagner and Zandi}(2014)}]{PhysRevA.90.012516}
\bibinfo{author}{\bibfnamefont{J.}~\bibnamefont{Wagner}} \bibnamefont{and}
  \bibinfo{author}{\bibfnamefont{R.}~\bibnamefont{Zandi}},
  \bibinfo{journal}{Phys. Rev.} \textbf{\bibinfo{volume}{A90}},
  \bibinfo{pages}{012516} (\bibinfo{year}{2014}).

\bibitem[{\citenamefont{Rodrigues et~al.}(2006)\citenamefont{Rodrigues,
  Maia~Neto, Lambrecht, and Reynaud}}]{Rodrigues:2006ku}
\bibinfo{author}{\bibfnamefont{R.~B.} \bibnamefont{Rodrigues}},
  \bibinfo{author}{\bibfnamefont{P.~A.} \bibnamefont{Maia~Neto}},
  \bibinfo{author}{\bibfnamefont{A.}~\bibnamefont{Lambrecht}},
  \bibnamefont{and} \bibinfo{author}{\bibfnamefont{S.}~\bibnamefont{Reynaud}},
  \bibinfo{journal}{Phys. Rev. Lett.} \textbf{\bibinfo{volume}{96}},
  \bibinfo{pages}{100402} (\bibinfo{year}{2006}).

\bibitem[{\citenamefont{Rodrigues et~al.}(2007)\citenamefont{Rodrigues,
  Maia~Neto, Lambrecht, and Reynaud}}]{Rodrigues:2007zza}
\bibinfo{author}{\bibfnamefont{R.}~\bibnamefont{Rodrigues}},
  \bibinfo{author}{\bibfnamefont{P.}~\bibnamefont{Maia~Neto}},
  \bibinfo{author}{\bibfnamefont{A.}~\bibnamefont{Lambrecht}},
  \bibnamefont{and} \bibinfo{author}{\bibfnamefont{S.}~\bibnamefont{Reynaud}},
  \bibinfo{journal}{Phys. Rev.} \textbf{\bibinfo{volume}{A75}},
  \bibinfo{pages}{062108} (\bibinfo{year}{2007}).

\bibitem[{\citenamefont{Chen et~al.}(2007)\citenamefont{Chen, Mohideen,
  Klimchitskaya, and Mostepanenko}}]{Chen:2007jea}
\bibinfo{author}{\bibfnamefont{F.}~\bibnamefont{Chen}},
  \bibinfo{author}{\bibfnamefont{U.}~\bibnamefont{Mohideen}},
  \bibinfo{author}{\bibfnamefont{G.}~\bibnamefont{Klimchitskaya}},
  \bibnamefont{and}
  \bibinfo{author}{\bibfnamefont{V.}~\bibnamefont{Mostepanenko}},
  \bibinfo{journal}{Phys. Rev. Lett.} \textbf{\bibinfo{volume}{98}},
  \bibinfo{pages}{068901} (\bibinfo{year}{2007}).

\bibitem[{\citenamefont{Rayleigh}(1907)}]{Rayleigh07}
\bibinfo{author}{\bibfnamefont{L.}~\bibnamefont{Rayleigh}},
  \bibinfo{journal}{Proceedings of the Royal Society of London. Series A}
  \textbf{\bibinfo{volume}{79}}, \bibinfo{pages}{399} (\bibinfo{year}{1907}).

\bibitem[{\citenamefont{Lippmann}(1953)}]{Lippman53}
\bibinfo{author}{\bibfnamefont{B.~A.} \bibnamefont{Lippmann}},
  \bibinfo{journal}{J. Opt. Soc. Am.} \textbf{\bibinfo{volume}{43}},
  \bibinfo{pages}{408} (\bibinfo{year}{1953}).

\bibitem[{\citenamefont{Uretsky}(1965)}]{Uretsky65}
\bibinfo{author}{\bibfnamefont{J.~L.} \bibnamefont{Uretsky}},
  \bibinfo{journal}{Annals of Physics} \textbf{\bibinfo{volume}{33}},
  \bibinfo{pages}{400} (\bibinfo{year}{1965}).

\bibitem[{\citenamefont{Millar}(1969)}]{Millar69}
\bibinfo{author}{\bibfnamefont{R.~F.} \bibnamefont{Millar}},
  \bibinfo{journal}{Mathematical Proceedings of the Cambridge Philosophical
  Society} \textbf{\bibinfo{volume}{65}}, \bibinfo{pages}{773}
  (\bibinfo{year}{1969}), ISSN \bibinfo{issn}{1469-8064}.

\bibitem[{\citenamefont{Millar}(1971)}]{Millar71}
\bibinfo{author}{\bibfnamefont{R.~F.} \bibnamefont{Millar}},
  \bibinfo{journal}{Mathematical Proceedings of the Cambridge Philosophical
  Society} \textbf{\bibinfo{volume}{69}}, \bibinfo{pages}{217}
  (\bibinfo{year}{1971}), ISSN \bibinfo{issn}{1469-8064}.

\bibitem[{\citenamefont{Waterman}(1974)}]{Waterman74}
\bibinfo{author}{\bibfnamefont{P.~C.} \bibnamefont{Waterman}},
  \bibinfo{journal}{Journal of the Acoustical Society of America}
  \textbf{\bibinfo{volume}{57}}, \bibinfo{pages}{791} (\bibinfo{year}{1974}).

\bibitem[{\citenamefont{Tatarskii}(1995)}]{Tatarskii95}
\bibinfo{author}{\bibfnamefont{V.~I.} \bibnamefont{Tatarskii}},
  \bibinfo{journal}{J. Opt. Soc. Am.} \textbf{\bibinfo{volume}{A12}},
  \bibinfo{pages}{1254} (\bibinfo{year}{1995}).

\bibitem[{\citenamefont{{Emig}}(2003)}]{Emig2003}
\bibinfo{author}{\bibfnamefont{T.}~\bibnamefont{{Emig}}},
  \bibinfo{journal}{Europhys. Lett.} \textbf{\bibinfo{volume}{62}},
  \bibinfo{pages}{466} (\bibinfo{year}{2003}).

\bibitem[{\citenamefont{Buscher and Emig}(2004)}]{Buscher:2004tb}
\bibinfo{author}{\bibfnamefont{R.}~\bibnamefont{Buscher}} \bibnamefont{and}
  \bibinfo{author}{\bibfnamefont{T.}~\bibnamefont{Emig}},
  \bibinfo{journal}{Phys. Rev.} \textbf{\bibinfo{volume}{A69}},
  \bibinfo{pages}{062101} (\bibinfo{year}{2004}).

\bibitem[{\citenamefont{Kats}(1977)}]{Kats77}
\bibinfo{author}{\bibfnamefont{E.~I.} \bibnamefont{Kats}},
  \bibinfo{journal}{Sov. Phys. JETP} \textbf{\bibinfo{volume}{46}},
  \bibinfo{pages}{109} (\bibinfo{year}{1977}).

\bibitem[{\citenamefont{Jaekel and Reynaud}(1991)}]{Jaekel91}
\bibinfo{author}{\bibfnamefont{M.~T.} \bibnamefont{Jaekel}} \bibnamefont{and}
  \bibinfo{author}{\bibfnamefont{S.}~\bibnamefont{Reynaud}},
  \bibinfo{journal}{J. Physique I} \textbf{\bibinfo{volume}{1}},
  \bibinfo{pages}{1395} (\bibinfo{year}{1991}).

\bibitem[{\citenamefont{Bulgac et~al.}(2006)\citenamefont{Bulgac, Magierski,
  and Wirzba}}]{Bulgac06}
\bibinfo{author}{\bibfnamefont{A.}~\bibnamefont{Bulgac}},
  \bibinfo{author}{\bibfnamefont{P.}~\bibnamefont{Magierski}},
  \bibnamefont{and} \bibinfo{author}{\bibfnamefont{A.}~\bibnamefont{Wirzba}},
  \bibinfo{journal}{Phys. Rev.} \textbf{\bibinfo{volume}{D73}},
  \bibinfo{pages}{025007} (\bibinfo{year}{2006}).

\bibitem[{\citenamefont{Lambrecht et~al.}(2006)\citenamefont{Lambrecht, {Maia
  Neto}, and Reynaud}}]{Lambrecht06}
\bibinfo{author}{\bibfnamefont{A.}~\bibnamefont{Lambrecht}},
  \bibinfo{author}{\bibfnamefont{P.~A.} \bibnamefont{{Maia Neto}}},
  \bibnamefont{and} \bibinfo{author}{\bibfnamefont{S.}~\bibnamefont{Reynaud}},
  \bibinfo{journal}{New J. Phys.} \textbf{\bibinfo{volume}{8}},
  \bibinfo{pages}{243} (\bibinfo{year}{2006}).

\bibitem[{\citenamefont{Kenneth and Klich}(2006)}]{Kenneth06}
\bibinfo{author}{\bibfnamefont{O.}~\bibnamefont{Kenneth}} \bibnamefont{and}
  \bibinfo{author}{\bibfnamefont{I.}~\bibnamefont{Klich}},
  \bibinfo{journal}{Phys. Rev. Lett.} \textbf{\bibinfo{volume}{97}},
  \bibinfo{eid}{160401} (\bibinfo{year}{2006}).

\bibitem[{\citenamefont{Emig et~al.}(2007)\citenamefont{Emig, Graham, Jaffe,
  and Kardar}}]{spheres}
\bibinfo{author}{\bibfnamefont{T.}~\bibnamefont{Emig}},
  \bibinfo{author}{\bibfnamefont{N.}~\bibnamefont{Graham}},
  \bibinfo{author}{\bibfnamefont{R.~L.} \bibnamefont{Jaffe}}, \bibnamefont{and}
  \bibinfo{author}{\bibfnamefont{M.}~\bibnamefont{Kardar}},
  \bibinfo{journal}{Phys. Rev. Lett.} \textbf{\bibinfo{volume}{99}},
  \bibinfo{pages}{170403} (\bibinfo{year}{2007}).

\bibitem[{\citenamefont{Kenneth and Klich}(2008)}]{Kenneth08}
\bibinfo{author}{\bibfnamefont{O.}~\bibnamefont{Kenneth}} \bibnamefont{and}
  \bibinfo{author}{\bibfnamefont{I.}~\bibnamefont{Klich}},
  \bibinfo{journal}{Phys. Rev. B} \textbf{\bibinfo{volume}{78}},
  \bibinfo{eid}{014103} (\bibinfo{year}{2008}).

\bibitem[{\citenamefont{Rahi et~al.}(2009)\citenamefont{Rahi, Emig, Graham,
  Jaffe, and Kardar}}]{universal}
\bibinfo{author}{\bibfnamefont{S.~J.} \bibnamefont{Rahi}},
  \bibinfo{author}{\bibfnamefont{T.}~\bibnamefont{Emig}},
  \bibinfo{author}{\bibfnamefont{N.}~\bibnamefont{Graham}},
  \bibinfo{author}{\bibfnamefont{R.~L.} \bibnamefont{Jaffe}}, \bibnamefont{and}
  \bibinfo{author}{\bibfnamefont{M.}~\bibnamefont{Kardar}},
  \bibinfo{journal}{Phys. Rev.} \textbf{\bibinfo{volume}{D80}},
  \bibinfo{pages}{085021} (\bibinfo{year}{2009}).

\bibitem[{\citenamefont{Calogero}(1967)}]{variable}
\bibinfo{author}{\bibfnamefont{F.}~\bibnamefont{Calogero}},
  \emph{\bibinfo{title}{Variable Phase Approach to Potential Scattering}}
  (\bibinfo{publisher}{Academic Press}, \bibinfo{address}{New York},
  \bibinfo{year}{1967}).

\bibitem[{\citenamefont{Graham et~al.}(2009)\citenamefont{Graham, Quandt, and
  Weigel}}]{Graham09}
\bibinfo{author}{\bibfnamefont{N.}~\bibnamefont{Graham}},
  \bibinfo{author}{\bibfnamefont{M.}~\bibnamefont{Quandt}}, \bibnamefont{and}
  \bibinfo{author}{\bibfnamefont{H.}~\bibnamefont{Weigel}},
  \emph{\bibinfo{title}{Spectral Methods in Quantum Field Theory}}
  (\bibinfo{publisher}{Springer-Verlag}, \bibinfo{address}{Berlin},
  \bibinfo{year}{2009}).

\bibitem[{\citenamefont{Forrow and Graham}(2012)}]{Forrow:2012sp}
\bibinfo{author}{\bibfnamefont{A.}~\bibnamefont{Forrow}} \bibnamefont{and}
  \bibinfo{author}{\bibfnamefont{N.}~\bibnamefont{Graham}},
  \bibinfo{journal}{Phys. Rev.} \textbf{\bibinfo{volume}{A86}},
  \bibinfo{pages}{062715} (\bibinfo{year}{2012}).

\bibitem[{\citenamefont{Newton}(1966)}]{Newton02}
\bibinfo{author}{\bibfnamefont{R.~G.} \bibnamefont{Newton}},
  \emph{\bibinfo{title}{Scattering Theory of Waves and Particles}}
  (\bibinfo{publisher}{McGraw-Hill}, \bibinfo{address}{New York},
  \bibinfo{year}{1966}).

\bibitem[{\citenamefont{Graham et~al.}(2002)\citenamefont{Graham, Jaffe,
  Khemani, Quandt, Scandurra, and Weigel}}]{density}
\bibinfo{author}{\bibfnamefont{N.}~\bibnamefont{Graham}},
  \bibinfo{author}{\bibfnamefont{R.~L.} \bibnamefont{Jaffe}},
  \bibinfo{author}{\bibfnamefont{V.}~\bibnamefont{Khemani}},
  \bibinfo{author}{\bibfnamefont{M.}~\bibnamefont{Quandt}},
  \bibinfo{author}{\bibfnamefont{M.}~\bibnamefont{Scandurra}},
  \bibnamefont{and} \bibinfo{author}{\bibfnamefont{H.}~\bibnamefont{Weigel}},
  \bibinfo{journal}{Nucl. Phys.} \textbf{\bibinfo{volume}{B645}},
  \bibinfo{pages}{49} (\bibinfo{year}{2002}).

\bibitem[{\citenamefont{Dunham and Graham}()}]{NGJSD}
\bibinfo{author}{\bibfnamefont{J.~S.} \bibnamefont{Dunham}} \bibnamefont{and}
  \bibinfo{author}{\bibfnamefont{N.}~\bibnamefont{Graham}}, \bibinfo{note}{work
  in progress}.

\end{thebibliography}

\end{document}